\documentclass[twocolumn]{aastex63}

\usepackage{gensymb}
\usepackage[figuresright]{rotating}

\graphicspath{{./}{figures/}}

\begin{document}

\title{Deep Search for Molecular Oxygen in TW Hya}


%
\correspondingauthor{Becky J. Williams}
\email{rjw9dmj@virginia.edu}

\author[0000-0002-1548-6811]{\textbf{Becky J. Williams}}
\affiliation{Department of Astronomy, University of Virginia, Charlottesville, VA 22904, USA}

\author[0000-0003-2076-8001]{\textbf{L. Ilsedore Cleeves}}
\affiliation{Department of Astronomy, University of Virginia, Charlottesville, VA 22904, USA}

\author[0000-0002-8743-1318]{\textbf{Christian Eistrup}}
\affiliation{Max Planck Institute for Astronomy, Königstuhl 17, 69117 Heidelberg, Germany}
\affiliation{Department of Astronomy, University of Virginia, Charlottesville, VA 22904, USA}

\author[0000-0002-3835-3990]{\textbf{Jon P. Ramsey}}
\affiliation{Department of Astronomy, University of Virginia, Charlottesville, VA 22904, USA}



\begin{abstract}

The dominant form of oxygen in cold molecular clouds is gas-phase carbon monoxide (CO) and ice-phase water (H$_2$O). Yet, in planet-forming disks around young stars, gas-phase CO and H$_2$O are less abundant relative to their ISM values, and no other major oxygen-carrying molecules have been detected. Some astrochemical models predict that gas-phase molecular oxygen (O$_2$) should be a major carrier of volatile oxygen in disks. We report a deep search for emission from the isotopologue $^{16}$O$^{18}$O ($N_J=2_1-0_1$ line at 233.946 GHz) in the nearby protoplanetary disk around TW Hya. We used imaging techniques and matched filtering to search for weak emission but do not detect $^{16}$O$^{18}$O. Based on our results, we calculate upper limits on the gas-phase O$_2$ abundance in TW Hya of $(6.4-70)\times10^{-7}$ relative to H, which is $2-3$ orders of magnitude below solar oxygen abundance. We conclude that gas-phase O$_2$ is not a major oxygen-carrier in TW Hya. Two other potential oxygen-carrying molecules, SO and SO$_2$, were covered in our observations, which we also do not detect. Additionally, we report a serendipitous detection of the C$^{15}$N $N_J = 2_{5/2}-1_{3/2}$ hyperfine transitions, $F = 3 - 2$ and $F = 2 - 1$, at 219.9 GHz, which we found via matched filtering and confirm through imaging.

\end{abstract}



\section{Introduction} \label{sec:intro}

Protoplanetary disks provide a critical link in understanding the chemical evolution from the interstellar medium (ISM) to planetary systems (including our own solar system). By studying these environments, we can understand how molecular abundances change as stars, and later planets, form. Such observations are helpful to put our own solar system into context. So far, there have been about 300 unique molecules (not including isotopologues) detected in the ISM. Within protoplanetary disks, only 25 unique molecules have been detected \citep{McGuire2022}. This low detection rate is not due to decreased chemistry in disks, but because we do not know some of the key tracers of the most abundant elements, such as oxygen.

Oxygen is the third most abundant element in the Universe, with a solar oxygen abundance of $4.9 \times 10^{-4}$ relative to hydrogen \citep{Asplund2009,Asplund2021}. The vast majority of oxygen in the Universe is incorporated into gas-phase molecules, ice-phase molecules, or refractory dust. \citet{Whittet2010} considered the incorporation of oxygen into silicates and oxides and estimated that the amount of oxygen in dust is $(0.9-1.4)\times10^{-4}$ relative to hydrogen in different ISM environments.

\citet{vanDishoeck2021} discuss the oxygen budget of ISM environments, assuming a total oxygen abundance of $5.8 \times 10^{-4}$ relative to hydrogen \citep{Przybilla2008}. They note that about $20\%$ of the total abundance of oxygen is unaccounted for in diffuse clouds, increasing to $50\%$ in dense regions. Two of the most common molecules in the ISM and protoplanetary disks are carbon monoxide (CO) and water (H$_2$O). In warm gas in the interstellar medium, gas-phase CO has an abundance of about $10^{-4}$ relative to hydrogen \citep{Ripple2013}, but is observed to be less abundant in protoplanetary disks \citep{Dutrey1994,Ansdell2016,Long2017}. Similarly, water ice has an abundance of H$_2$O/H$_2$ $\approx 10^{-4}$ in dense clouds \citep[see][]{vanDishoeck2013}, but water vapor has been detected in low abundance in only a few disks \citep{Du2017}. It is possible that a large amount of oxygen is in frozen species, such as water ice, that are difficult to observe. Another possibility is that oxygen is in other gas-phase molecules, such as molecular oxygen (O$_2$).

O$_2$ has been observed in low abundances in two molecular clouds: O$_2$/H$_2$ $\approx (0.3-7.3) \times 10^{-6}$ in Orion \citep{Goldsmith2011} and O$_2$/H$_2$ $\approx 5 \times 10^{-8}$ in $\rho$ Oph A \citep{Liseau2012}. O$_2$ is a reactive molecule, and its most common form, $^{16}$O$^{16}$O, is difficult to detect due to a lack of a permanent dipole moment. The less-abundant isotopologue $^{16}$O$^{18}$O does, however, have a dipole moment. Through this isotopologue, O$_2$ was tentatively detected in the protostellar system IRAS 16293-2422 B \citep{Taquet2018}. There have been no other reported detections of O$_2$ in protoplanetary disks.

Interestingly, O$_2$ has been detected in our own solar system in comets, which we do not expect to have undergone much chemical evolution since their formation in the pre-solar nebula. For example, O$_2$ was detected in the coma of comet 67P/Churyumov-Gerasimenko (67P) by the ROSINA \citep{Balsiger2007} instrument on ESA's Rosetta spacecraft. \citet{Bieler2015} found an unexpectedly high average O$_2$ to water ratio of $3.80 \pm 0.85\%$, making O$_2$ the fourth most abundant species in the coma. O$_2$ was also detected in comet 1P/Halley at an abundance of $3.7 \pm 1.7\%$ relative to water \citep{Rubin2015}. In both cases, O$_2$ and H$_2$O abundance appear to be correlated. As described in \citet[][and references therein]{LuspayKuti2018}, there have been many proposed origins for O$_2$ in comets, ranging from \textit{in situ} processes to origins in the ISM prior to formation of the solar nebula. Recent analysis by \citet{LuspayKuti2022} shows that, farther from the Sun, O$_2$ abundance in 67P is more strongly correlated with CO and CO$_2$ than with H$_2$O. They suggest that 67P has two reservoirs of O$_2$: a deep primordial nucleus of CO and CO$_2$ ice, and a surface layer of H$_2$O ice that formed later.

If O$_2$ in comets is not formed \textit{in situ}, then it must have been present in the protoplanetary disk from whence the comet formed. This scenario is supported by chemical models that predict a large amount of oxygen is in gas-phase O$_2$ in the inner regions of disks. For example, O$_2$ ice can be produced on dust grain surfaces and under certain conditions desorbs faster than it reacts to form other molecules. \citet{Eistrup2018} predict that this process occurs in the midplane of disks, resulting in a build-up of gas-phase O$_2$ between the H$_2$O and O$_2$ ice lines (between 0.7 and 10 au after 8 million years). \citet{Walsh2015} meanwhile predict that gas-phase O$_2$ builds up in the atmosphere of T Tauri disks, produced via gas-phase neutral-neutral reactions of O $+$ OH. They predict that O$_2$ may carry 50$\%$ of the total oxygen in the disk's atmosphere, and 10$\%$ when including the midplane.

In this paper, we present a deep search for $^{16}$O$^{18}$O emission in ALMA observations of TW Hya, a well-studied T Tauri protoplanetary disk. TW Hya's relatively gas-rich disk and nearby proximity make it a good candidate for searching for faint emission. There have been many observations of gas emission lines in TW Hya, and previous observations have shown that CO gas is 1-2 orders of magnitude less abundant than in the ISM \citep{Favre2013,Cleeves2015co}. Water vapor has a maximum observed abundance of H$_2$O/H$_2$ $\approx10^{-7}$ in TW Hya \citep{Hogerheijde2011}. Evidently, gas-phase CO and H$_2$O do not account for the majority of oxygen in TW Hya. Could oxygen be hiding in gas-phase O$_2$?

\section{Observations} \label{sec:obs}

Observations of the $^{16}$O$^{18}$O $N_J=2_1-0_1$ line at 233.946 GHz toward TW Hya were carried out as part of ALMA program 2019.1.01177.S (PI: Eistrup).  Observations occurred on two nights in 2021, April 4 and April 6, for a combined time of 91 minutes on source. On April 4, there were 44 antennas and baselines of 15-1398 meters, and on April 6, 45 antennas and baselines of 15-1263 meters. J1058+0133 was adopted as the amplitude and bandpass calibrator, while J1037-2934 was used for phase calibration. The data were calibrated using the standard ALMA pipeline.

We searched for $^{16}$O$^{18}$O emission in both the image plane and the visibility plane, as described in more detail in the following section. Subsequent imaging and analysis were carried out using the Common Astronomy Software Applications (CASA) version 6.4 \citep{CASA2007}. Continuum emission was subtracted using the task \textit{uvcontsub}, applying a fit order of 1 and excluding edge channels. The data were imaged using \textit{tclean}, initially with a natural weighting scheme to improve line sensitivity. The resulting beam is $0.59''$ by $ 0.50''$ with a position angle of $-88\degree$, equivalent to a spatial resolution of 30 to 35 au at a distance of 60.1 pc \citep{BailerJones2018}. The rms noise is 2.59 mJy beam$^{-1}$ for a channel width of 0.235 km s$^{-1}$. For the matched filtering analysis, Earth's rotation was first corrected for using the CASA task \textit{cvel}.

\section{Methods} \label{sec:methods}

We first created a Keplerian mask \citep{TeagueKepMask} to use with the CASA \textit{tclean} task. Due to the rotation of disks, we observe some emission to be shifted to higher or lower frequencies, so certain regions of the disk will be brighter in different channels. A Keplerian mask traces this emission pattern, based on the parameters of the disk, to improve the signal to noise of the integrated intensity \citep[e.g.][]{Salinas2017,Teague2022}. We assumed a systemic velocity of 2.86 km s$^{-1}$ \citep{Favre2013}, distance of 60.1 pc \citep{BailerJones2018}, inclination $= 5.8\degree$, position angle $=151\degree$, and stellar mass $= 0.81$ M$_\odot$ \citep{Teague2019}. The mask was created using CO $J=3-2$ emission data for TW Hya from \citet{Huang2018}, and parameters optimized to fully trace the CO emission. Specifically, we set the \textit{target\_res} = 0.4, which convolves the mask with a circular beam of 0.4 arcsec FWHM, and \textit{dV0} = 500 m s$^{-1}$, which affects the radially varying line width of emission. Two masks were then created for the O$_2$ data, one with a radius set to 100 au (convolved to 120 au), and a second with a radius set to 240 au (convolved to 260 au, covering the extent of the CO emission). Both masks were used separately in cleaning, using a $5\sigma$ threshold. The resulting image was the same in both cases because there were no values above $5\sigma$ within either mask.

The \textit{tclean} task has several parameters that can be varied to bring out faint emission.  For example, the choice of weighting scheme and uvtaper can be tuned to give a higher weighting to shorter baselines to increase sensitivity, at the expense of a lower resolution. Because we were mainly interested in looking at the disk integrated flux, we did not need the full spatial resolution of the observations. We imaged the data with several combinations of the Briggs weighting robust parameter \citep{Briggs1995} and uvtaper. We used channel averaging as another method of revealing faint emission. The original channel width of the data was 0.078 km s$^{-1}$, and we experimented with averaging 2, 3, and 5 channels together.

\begin{figure*}
\begin{center}
\includegraphics[width=.99\textwidth]{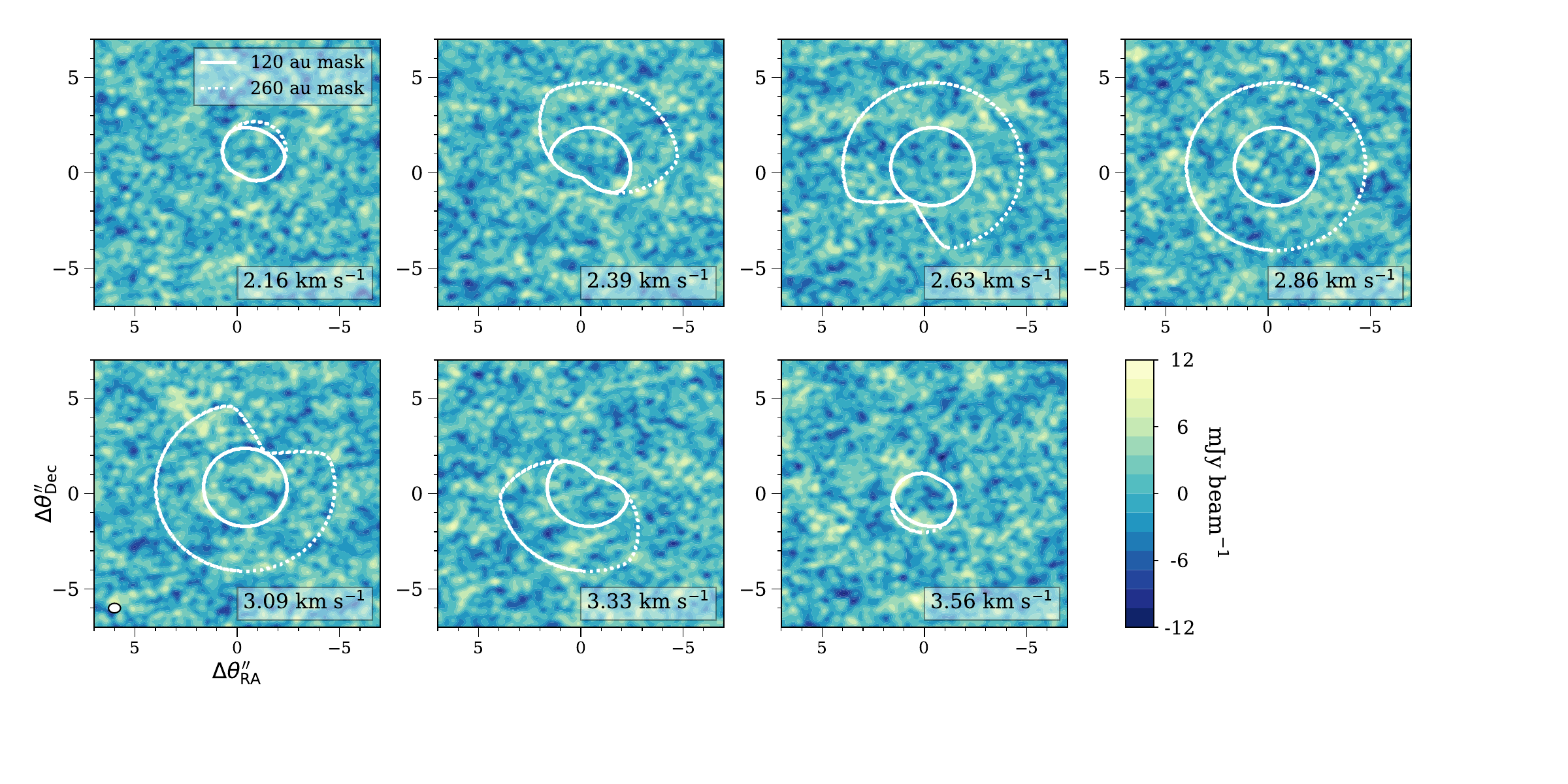}
\caption{Channel maps of the intensity, centered on 233.946 GHz. The systemic velocity is 2.86 km s$^{-1}$. The two Keplerian masks are shown in white contours, and velocities are shown in the lower right of each panel. The beam size is $0.59'' \times 0.50''$ and is shown in the lower left panel.}
\label{fig:channels}
\end{center}
\end{figure*}

Matched filtering is a technique to detect weak emission in observational data \citep{Loomis2018}. Matched filtering is applied in the visibility plane and therefore does not involve any imaging. It requires a filter that models the expected pattern of emission across velocity channels. The filter is Fourier transformed to generate a kernel that is then cross-correlated with the data. The result is an impulse response spectrum, with peaks indicating possible emission. For matched filtering, we used both Keplerian masks as filters (one with a 120 au radius, and one with a 260 au radius). It is possible that O$_2$ emission covers a smaller area than these radii and could be hidden in the noise. However, Keplerian masks cover less area in edge channels, so a smaller mask is below the resolution of the images.

\section{Results} \label{sec:result}

\subsection{Imaging Results} \label{imresult}
\begin{figure}
\begin{center}
\includegraphics[width=.47\textwidth]{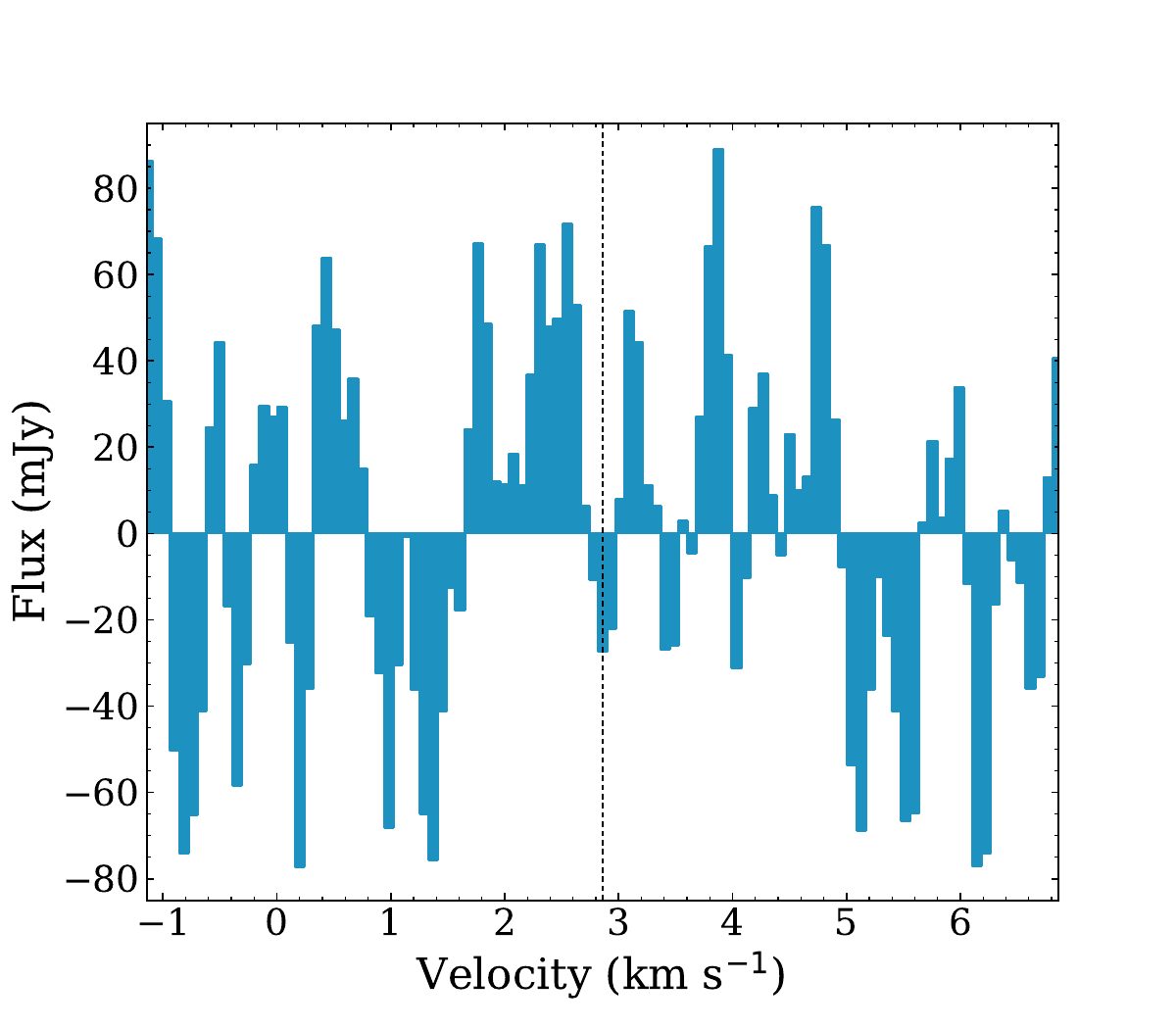}
\caption{Spectrum of the O$_2$ data for TW Hya, calculated within a 260 au region. The dashed line is the systemic velocity of 2.86 km s$^{-1}$, centered on 233.946 GHz.}
\label{fig:o2_spec}
\end{center}
\end{figure}

Figure~\ref{fig:channels} shows the data imaged with natural weighting and velocity averaging to a channel size of 0.235 km s$^{-1}$. For all attempted combinations of imaging parameters, no significant detection was found within either choice of mask. A spectrum, calculated within 260 au, is shown in Figure \ref{fig:o2_spec}. The average rms noise level per channel in Figure~\ref{fig:channels} is 2.59 mJy beam$^{-1}$. We calculated a disk integrated flux (for a velocity range of 4.69 km s$^{-1}$) of $-11.8 \pm 14.2$ mJy km s$^{-1}$ within 120 au, and $7.4 \pm 18.0$ mJy km s$^{-1}$ within 260 au; i.e., at the noise level of the data, we do not detect $^{16}$O$^{18}$O $N_J=2_1-0_1$ emission.

\subsection{Matched Filter Results} \label{matfilresult}

\begin{figure*}
\begin{center}
\includegraphics[width=1\textwidth]{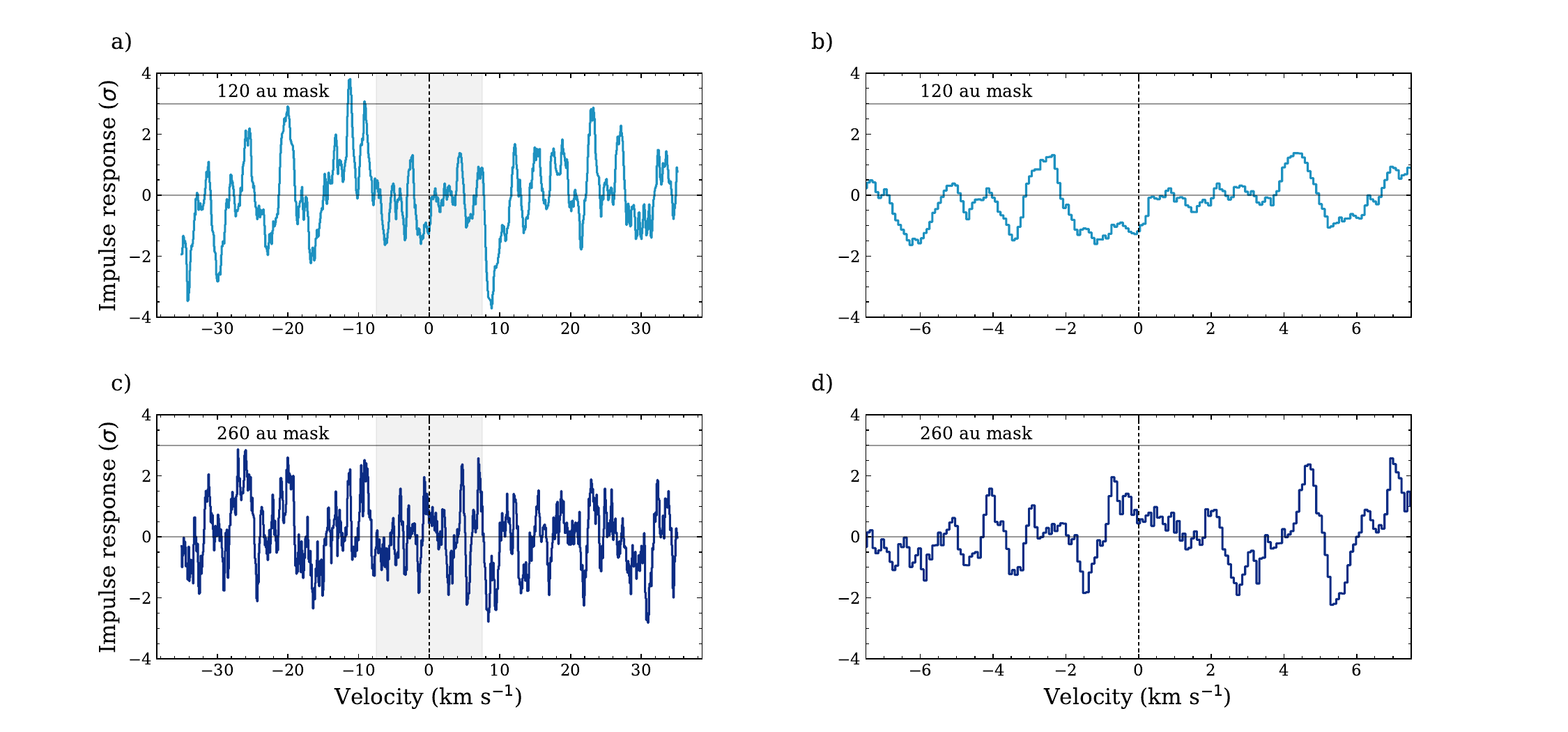}
\caption{Matched filtering response for the $^{16}$O$^{18}$O data, with impulse response on the y-axis and velocity on the x-axis. The 233.946 GHz transition is centered at 0 km s$^{-1}$ and is denoted by the vertical dashed line. The two horizontal lines are at 0$\sigma$ and 3$\sigma$. The gray region in the left plots is the range of velocities covered in the right plots. a) 120 au filter, covering all channels. b) zoom-in of the shaded region in panel a. c) 260 au filter, covering all channels. d) zoom-in of the shaded region in panel c.}
\label{fig:matfil}
\end{center}
\end{figure*}

While we cannot see emission in the image plane, if there is emission just below the noise level, matched filtering has previously been successful at finding the signatures of line emission in other sources \citep{Loomis2020}, including when using a Keplerian mask as a template \citep{Loomis2018}. By applying the Keplerian mask as our template, we provide an expectation for where rotating gas emission should appear spatially in our data. Since we do not know the extent of the O$_2$, we try both the 120 au and 260 au masks as templates. The results of matched filtering the data with these templates are presented in Figure \ref{fig:matfil}. A $3\sigma$ response at 0 km s$^{-1}$ would indicate emission from $^{16}$O$^{18}$O. Neither filter resulted in a $3\sigma$ impulse response. We can conclusively say that the $^{16}$O$^{18}$O 233.946 GHz line is not detected at the sensitivity of the present measurements.

\subsection{O$_2$ Upper Limits} \label{upperlim}

 How much oxygen could be hiding within our upper limit on $^{16}$O$^{18}$O emission? To estimate the total number of O$_2$ atoms from a single line, we explore a range of temperatures and assume its emission is in local thermodynamic equilibrium (LTE). Using the image-plane disk integrated flux limits from Section \ref{imresult}, we calculate the 1$\sigma$ upper limit on the maximum number of O$_2$ molecules that could be present in TW Hya, using the following adapted from \citet{Bergin2013}. 
 
 \begin{equation}\label{eq1}
 F_l = \frac{\mathcal{N}_{^{16}O^{18}O}A_{20}h \nu f_u}{4 \pi D^2}
 \end{equation}
 $F_l$ is the 1$\sigma$ flux calculated from our images, $\mathcal{N}_{^{16}O^{18}O}$ is the number of $^{16}$O$^{18}$O molecules, $A_{20}=1.33\times10^{-8}$ s$^{-1}$ is the Einstein A coefficient of the $N_J=2_1-0_1$ transition \citep{Marechal1997}, $\nu=233.94618$ GHz is the frequency of the $N_J=2_1-0_1$ transition, and $D=60.1$ pc is the distance to TW Hya \citep{BailerJones2018}. The fraction of molecules in the upper state is given by $f_u = 3.0 \exp(-11.23$ K$/T) / Q(T)$, where 3 is the upper state degeneracy, 11.23 K is the upper state energy \citep{Marechal1997}, $T_{\rm gas}$ is the gas temperature, and $Q(T)$ is the partition function from the JPL spectral line catalog \citep{Pickett1998,Mizushima1991,Crownover1990,Steinbach1975,Amano1974}. These partition functions are given for temperatures ranging from 9.4 K to 300.0 K, so this is the temperature range we use in our calculations. To get an upper limit on the total number of $^{16}$O$^{16}$O molecules, we assumed a $^{16}$O$^{16}$O/$^{16}$O$^{18}$O ratio of 280 \citep{Taquet2018,Wilson1994}. We obtained an upper limit of $(1.1-10.4)\times10^{49}$ O$_2$ molecules within 120 au, and $(1.4-13.2)\times10^{49}$ molecules within 260 au, for a temperature range of 9.4 K to 300.0 K.

To give our result context, we also estimate the disk-averaged abundance of O$_2$ relative to hydrogen. To estimate the total hydrogen mass in the disk, we adopt the disk mass reported in \citet{Calahan2021}, but we note that there are a wide range of mass values for the TW Hya disk in the literature \citep[see][]{Miotello2022}. We adopt the \citet{Calahan2021} value of $2.5\times10^{-2}$ M$_\odot$ since it uses the HD line and multiple CO isotopologues to constrain the temperature structure. Assuming a molecular mass per hydrogen molecule of 2.8 \citep{Kauffman2008}, this equates to $2.1\times10^{55}$ hydrogen atoms in the whole disk. Within 120 au, we use the surface density profile from \citet{Calahan2021} to calculate a disk mass of  $1.8\times10^{-2}$ M$_\odot$, which equates to $1.5\times10^{55}$ hydrogen atoms. Using these values, we estimate an O$_2$/H abundance of $(7.2-70)\times10^{-7}$ within 120 au, and $(6.4-62)\times10^{-7}$ within 260 au, for temperatures ranging from 9.4 K to 300.0 K. We will return to the implications of these results in Section \ref{sec:discussion}.

\subsection{Constraints from Serendipitous Molecular Lines} \label{otherobs}

The SO $N_J=5_6-4_5$ and SO$_2$ $N_J=4_{(2,2)}-3_{(1,3)}$ lines at 219.949 GHz and 235.152 GHz, respectively, were also included in the observational set-up. Given that these molecules might be prominent oxygen carriers, we searched for emission from these lines using similar techniques as for $^{16}$O$^{18}$O. Neither line was detected above $3\sigma$ with imaging or matched filtering (see Appendix \ref{appendix}). The disk integrated flux for the SO line is $-0.05 \pm 12.5$ mJy km s$^{-1}$ within 120 au and $-21.4 \pm 21.2$ mJy km s$^{-1}$ within 260 au (covering a range of 4.99 km s$^{-1}$). For SO$_2$, the flux is $24.4 \pm 14.8$ mJy km s$^{-1}$ within 120 au and $34.3 \pm 24.5$ mJy km s$^{-1}$ within 260 au (covering a range of 4.67 km s$^{-1}$). We modified Equation \ref{eq1} for SO and SO$_2$, using values from LAMDA \citep{Schoier2005} and CDMS \citep{Muller2001}, and partition functions from the JPL spectral line catalog \citep{Pickett1998,Amano1974,Clark1976,Helminger1985,Lovas1985,Alekseev1996}. We calculate a 1$\sigma$ upper limit of SO/H = $(2.4-10)\times10^{-13}$ within 120 au and $(2.9-12)\times10^{-13}$ within 260 au, and SO$_2$/H = $(7.9-197)\times10^{-13}$ within 120 au and $(9.1-229)\times10^{-13}$ within 260 au, for a temperature range of 9.4 K to 300.0 K.

\begin{figure}
\begin{center}
\includegraphics[width=.47\textwidth]{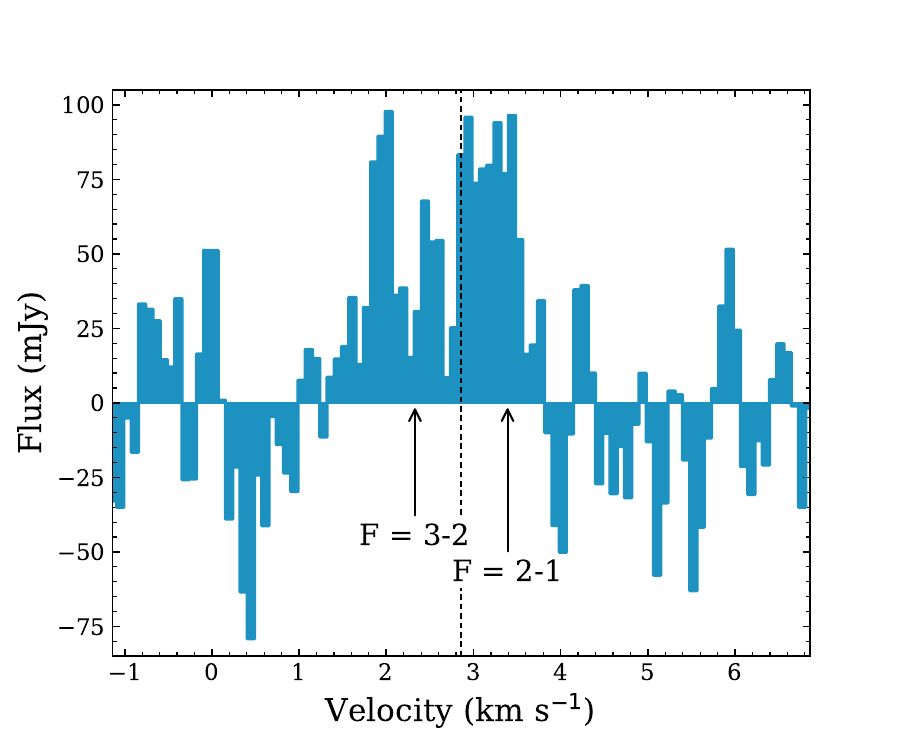}
\caption{Spectrum of the C$^{15}$N detection in TW Hya, calculated within a 200 au circular region. The two transitions are $N = 2-1, J = 5/2 - 3/2$ transitions, $F = 3 - 2$ and $F = 2 - 1$. The systemic velocity of 2.86 km s$^{-1}$, centered on the average of the two emission frequencies, is marked with the dashed line.}
\label{fig:c15n}
\end{center}
\end{figure}

We also report a serendipitous detection of the C$^{15}$N $N_J = 2_{5/2}-1_{3/2}$ hyperfine transitions, $F = 2 - 1$ and $F = 3 - 2$, at 219.93404 GHz and 219.93482 GHz, respectively. Using matched filtering with the 120 au Keplerian mask, we obtained a 7$\sigma$ filter response. The emission is visible in images (not shown), with an integrated flux of $113.5 \pm 17.0$ mJy km s$^{-1}$ for a range of 2.33 km s$^{-1}$, calculated within a 200 au circular region covering visible emission. For these images, we used Briggs weighting (robust = 2) and a uvtaper of 0.3 arcsec. The flux per channel is shown in Figure~\ref{fig:c15n}. C$^{15}$N has previously been detected in TW Hya through its $N = 3-2$ transition \citep{HilyBlant2017}. They used two methods to calculate an integrated flux of $150 \pm 20$ mJy km s$^{-1}$ and $160 \pm 13$ mJy km s$^{-1}$, the first of which is consistent with our value.

\section{Discussion} \label{sec:discussion}

\begin{figure}
\includegraphics[width=0.47\textwidth]{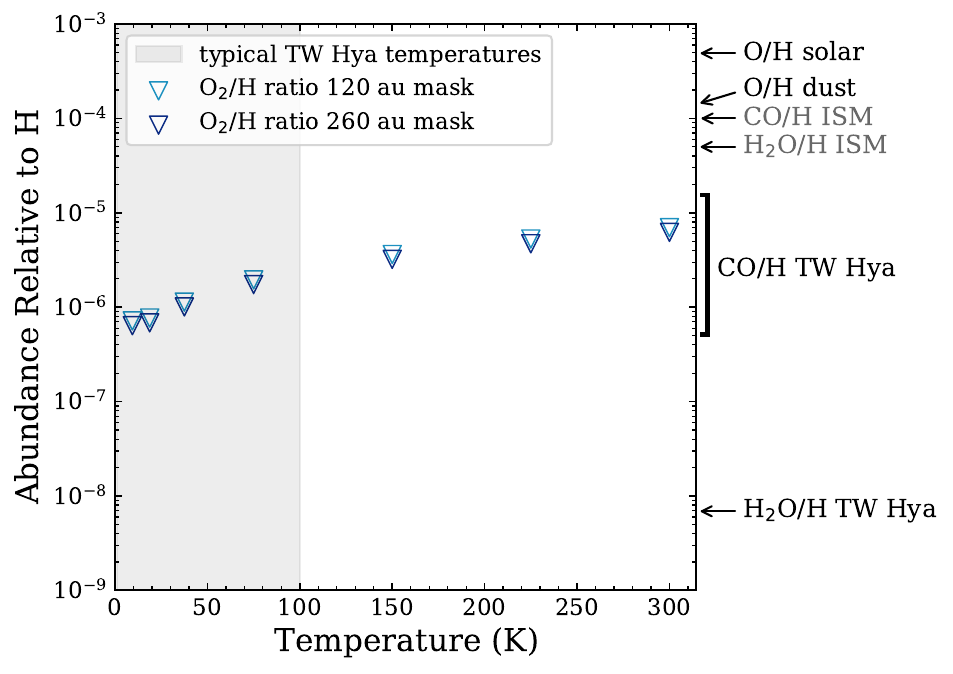}
\caption{Upper limits on O$_2$ in TW Hya as a function of temperature. The shaded region shows typical temperatures in TW Hya \citep[see][]{BerginCleeves2018}, the light blue and dark blue open triangles show the upper limits on O$_2$ abundance within 120 au and 260 au, respectively. Values for solar oxygen abundance \citep{Asplund2009} and other known oxygen-carrying species are shown along the right axis. For dust, we assumed a value of $1.4\times10^{-4}$ from \citet{Whittet2010}. From top to bottom, the remaining values used are from \citet{Ripple2013,vanDishoeck2013,Favre2013,Zhang2013}.}
\label{fig:uplim}
\end{figure}

How does O$_2$ stack up compared to other known oxygen carriers? Figure \ref{fig:uplim} shows our upper limits on O$_2$ abundance as a function of temperature, compared to the solar oxygen abundance and known abundances of other major oxygen-carrying species in the ISM and in TW Hya. Since CO and H$_2$O each contain one oxygen atom, their abundances relative to hydrogen can be directly compared to the solar oxygen abundance (e.g., if all oxygen were in CO, then the CO/H abundance would equal the O/H value). O$_2$, on the other hand, contains two oxygen atoms, so should be multiplied by two to compare with the solar oxygen abundance.

In TW Hya, both CO gas and H$_2$O gas are several orders of magnitude below solar oxygen abundance; neither molecule is a major reservoir of oxygen. Our upper limit shows that gas-phase O$_2$ is also not a major carrier of oxygen. Other oxygen-carrying molecules that have been detected in TW Hya include HCO$^{+}$ \citep{vanDishoeck2003}, H$_2$CO \citep{Oberg2017}, CH$_3$OH \citep{Walsh2016}, and HCOOH \citep{Favre2018}, but these molecules were all detected at too low an abundance to complete the oxygen budget relative to solar values. The majority of oxygen in TW Hya has not been detected, leading to several possibilities, which we discuss further.

Our upper limits on gas-phase O$_2$ are not constraining enough to rule out protoplanetary disks as the origin of O$_2$ in comets, as water vapor in TW Hya is detected in low abundance \citep{Zhang2013}. Comets 67P and 1P/Halley were observed to have high O$_2$/H$_2$O ratios of about 0.04 \citep{Bieler2015,Rubin2015}, whereas our upper limits on gas-phase O$_2$ are about two orders of magnitude more abundant than detected water vapor in TW Hya (see Figure \ref{fig:uplim}).

One possibility for the low detections of gas-phase oxygen carriers is that oxygen is frozen out, so it cannot be detected easily. Due to the low temperatures of disks, we expect many molecules (like H$_2$O) to exist mostly in the ice-phase, but several processes can return molecules to the gas-phase throughout the entire disk. Photodesorption occurs when ultraviolet photons strike a molecule on the surface of a grain, causing the molecule to break off from the grain. It depends on the photon flux, as well as the binding energy of the molecule. \citet{Oberg2009} studied the photodesorption yield of H$_2$O and found that the main products are H$_2$O and OH. They also found that at high temperatures (100 K), up to $20\%$ of the ice desorbs as O$_2$. \citet{Du2017} searched for cold water vapor in 13 protoplanetary disks, only detecting it in low abundance in two disks, and in the stacked spectrum of four other disks. They model the disk chemistry and find that, to match observations, the abundance of gas-phase oxygen must be reduced by a factor of 100 or more. They propose that oxygen (in the form of H$_2$O and CO) freezes onto dust grains, which then settles to the midplane \citep{Hogerheijde2011,Bergin2016}. This process primarily occurs in the outer disk; in the inner disk (within 15 au), temperatures are higher and frozen molecules may return to the gas phase. Grain size may also play a role in molecular abundances. \citet{Eistrup2022} modeled disk chemistry, taking into account grain size. They found that larger grain sizes result in a lower gas-phase O$_2$ abundance and a higher H$_2$O ice abundance, relative to the abundances produced using a fiducial 0.1$\mu$m grain size. As grain size increases, the surface area decreases, which decreases the number of grain surface-reactions and gas-grain interactions that can occur.

Another possibility is that there is a large amount of oxygen in gas-phase molecules that we have not yet observed. Overall, the most abundant molecules detected in comets are H$_2$O, CO$_2$, and CO \citep{Rubin2020}. As discussed already, H$_2$O and CO have been detected in disks in low abundance. Models by \citet{Eistrup2018} predict CO$_2$/H abundances of up to $\approx10^{-4}$ in disks. CO$_2$ gas is difficult to observe in disks because, like O$_2$, it is symmetric and lacks a permanent dipole moment. CO$_2$ has been detected in the disk within 3 au of AA Tauri \citep{Carr2008}, using Spitzer Space Telescope observations in the mid-infrared. JWST, however, can detect ice absorption features in the infrared, including CO$_2$ and H$_2$O ices, and is already providing insight into oxygen-carrying molecules in disks \citep[e.g.][]{Yang2022,Grant2023,McClure2023}.

\section{Conclusions} \label{sec:conclusions}

We searched for but did not detect emission from gas-phase $^{16}$O$^{18}$O in the protoplanetary disk around TW Hya. We used various imaging techniques along with matched filtering, and used our results to determine an upper limit on gas-phase O$_2$ in TW Hya.

\begin{itemize}
    \item The isotopologue $^{16}$O$^{18}$O was not detected in TW Hya, leading to an upper limit on the abundance of O$_2$ of $(7.2-70)\times10^{-7}$ relative to H if the emission is contained within 120 au. For the whole disk, the upper limit is $(6.4-62)\times10^{-7}$. This limit is 2-3 orders of magnitude lower than the solar oxygen abundance, so gas-phase O$_2$ is not a major reservoir for oxygen in TW Hya. Taking into account other existing molecular detections in TW Hya, the main oxygen-carrier(s) remain undetected.
    \item We place sensitive upper limits on the SO and SO$_2$ lines at 219.949 GHz and 235.152 GHz, respectively. We calculated an upper limit of SO/H = $(2.4-10)\times10^{-13}$ within 120 au and $(2.9-12)\times10^{-13}$ within 260 au, and of SO$_2$/H = $(7.9-197)\times10^{-13}$ within 120 au and $(9.1-229)\times10^{-13}$ within 260 au. These results suggest oxygen is not bound up with sulfur either.
    \item We detect the isotopologue C$^{15}$N at the 7$\sigma$ level using matched filtering, and calculate an integrated flux of $113.5 \pm 17.0$ mJy km s$^{-1}$.

\end{itemize}

It is difficult to determine the main reservoir of oxygen in disks because of the many solid and gaseous forms oxygen may take (e.g. frozen H$_2$O, CO, CO$_2$; gas-phase O$_2$, CO$_2$). More observations of disks are necessary to search for this missing reservoir. Focusing on TW Hya, future searches for other gas-phase molecules, such as isotopologues of CO$_2$, would provide more insight on oxygen. Alternatively, searches for gas-phase $^{16}$O$^{18}$O in other disks, could prove interesting. Observations of ices, especially with JWST, will also be helpful.

\vspace{9mm}
\acknowledgments 
B.J.W. acknowledges support from the Virginia Initiative on Cosmic Origins (VICO). L.I.C. acknowledges support from the David and Lucile Packard Foundation, Research Corporation for Science Advancement Cottrell Fellowship, NASA ATP 80NSSC20K0529, NSF grant no. AST-2205698, and SOFIA Award 09-0183. C.E. acknowledges support from VICO. J.P.R. acknowledges support from the NASA Astrophysics Theory Program under grant no.\ 80NSSC20K0533, from the National Science Foundation (NSF) under grant nos.\ AST-1910106 and AST-1910675, and from the Virginia Initiative on Cosmic Origins. This paper makes use of the following ALMA data: ADS/JAO.ALMA\#2019.1.01177.S. ALMA is a partnership of ESO (representing its member states), NSF (USA) and NINS (Japan), together with NRC (Canada) and NSC and ASIAA (Taiwan), in cooperation with the Republic of Chile. The Joint ALMA Observatory is operated by ESO, AUI/NRAO and NAOJ.

%

\facilities{ALMA}


\software{Astropy \citep{Astropy2013},  
          CASA \citep{CASA2007},
          Keplerian Mask \citep{TeagueKepMask}, VISIBLE \citep{LoomisVISIBLE}
          }



\vspace{9mm}

\appendix

\section{Matched Filter Results: Other Transitions}
\label{appendix}

\begin{figure*}[h]
\begin{center}
\includegraphics[width=1\textwidth]{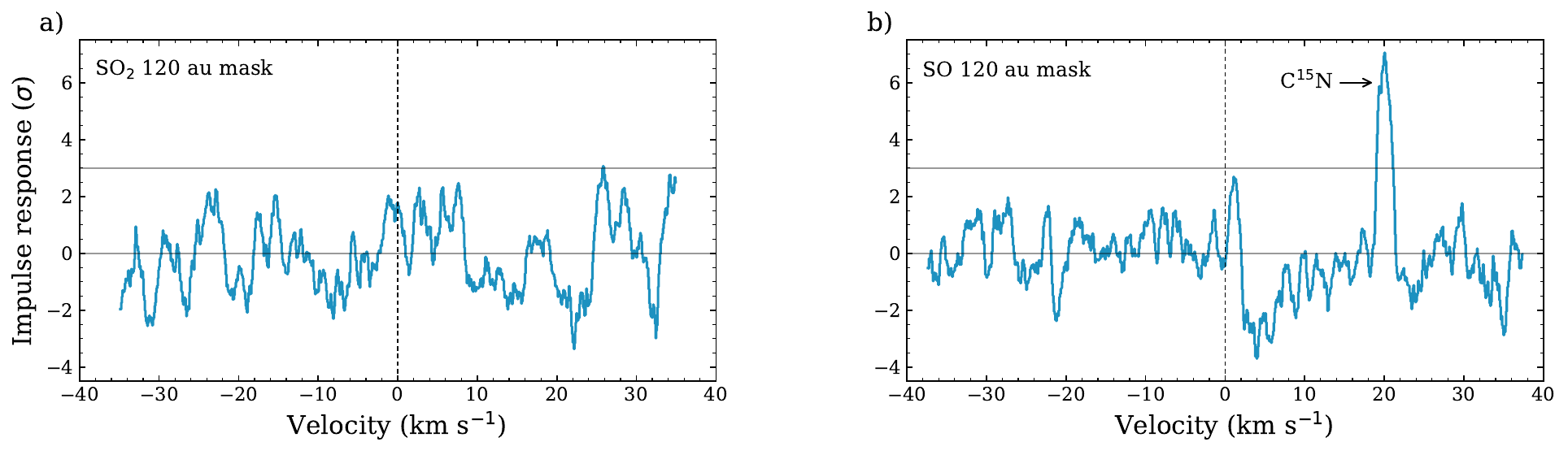}
\caption{Matched filtering response for the data containing the SO$_2$ and SO transitions, with impulse response on the y-axis and velocity on the x-axis. The two horizontal lines are at 0$\sigma$ and 3$\sigma$. a) SO$_2$ transition, centered at 235.152 GHz b) SO transition, centered at 219.949 GHz. The 7$\sigma$ peak is the C$^{15}$N detection.}
\label{fig:soso2}
\end{center}
\end{figure*}

The matched filtering response for the data containing the SO$_2$ and SO transitions is shown in Figure \ref{fig:soso2}. In both cases, the 120 au filter was used, and neither SO$_2$ nor SO were detected at the 3$\sigma$ level or above. The 7$\sigma$ peak at about 20 km s$^{-1}$ in the SO spectrum is the C$^{15}$N detection.

\end{document}